\documentstyle[12pt]{article}
\title{Two--poles $R$-matrices}
\author{Michel Talon\thanks{L.P.T.H.E. Universit\'e Paris VI
Bo\^{\i}te 126 /4 place Jussieu/ 75252 PARIS CEDEX 05}}
\date{\today}

\def\l{{\lambda}}
\def\G{{\cal G}}
\def\x {\stackrel {\textstyle \otimes}{,}}

\begin{document}
\bibliographystyle{perso}

\begin{titlepage}
\renewcommand{\thepage}{}
\maketitle
\vskip 2cm
\begin{abstract}
We study integrable dynamical systems described by a Lax pair
involving a spectral parameter. By solving the classical Yang--Baxter
equation when the $R$-matrix has two poles we show that they can be
interpreted as natural motions on a twisted loop algebra.
\end{abstract}
\vfill
{\sf Kingston July 1992}\hfill\break PAR-LPTHE 92/36 \hfill Work
supported by CNRS: URA 280
\end{titlepage}
\renewcommand{\thepage}{\arabic{page}}

\section{Introduction}

Let us consider some dynamical system whose equations of motion have
been written under a Lax form~\cite{Lax}:
\begin{equation}
{dL \over dt }= [L,M] \label{lax}
\end{equation}
Here, $L$ and $M$ belong to some Lie algebra $\G$, and we assume the
existence of globally defined maps (i.e, the Lax pair, $L$ and $M$)
from the phase space of our dynamical system to $\G$ such that the
equations of motion are equivalent to~(\ref{lax}).

Such a Lax formulation, when it exists, is generally not unique. In
particular there exists a general mechanism, known as dual moment
maps~\cite{AdHaHu}, allowing to switch between two different Lax
formulations of a dynamical system (frequently one with an $N\times N$
Lax pair, and the other one a $2\times 2$ Lax pair). Moreover, for
some systems, it is necessary in order to obtain a Lax formulation to
introduce an auxiliary parameter $\l$, called the spectral parameter,
and to consider a Lax pair $L(\l),\, M(\l)$ explicitely dependent on
the spectral parameter.  This means that the Lie algebra $\G$ in such
cases is a loop algebra ${\bf g}\otimes {\bf C}[\l,\l^{-1}]$ where
$\bf g$ is an ordinary matrix Lie algebra.

A nice example is provided by the Toda chain. Considering the open
chain as a Dynkin diagram naturally leads to a Lax formulation without
spectral parameter. The similar construction for the closed chain
leads to a loop algebra.  Moreover there exists an alternative
approach due to Sklyanin~\cite{Skl2} which builds a Lax pair by
considering the product of $2\times 2$ ``local'' monodromy matrices
depending on a spectral parameter. Several related examples are
similarly discussed in~\cite{graded,RSTS2}.

In spite of the arbitrariness involved in a Lax formulation, it is a
good first step towards the solution of the dynamical problem, since
it immediately allows to find conserved quantities. As a matter of
fact, eq.~(\ref{lax}) implies conservation of the spectrum of $L$, as
first emphasized by Lax~\cite{Lax}. In other words, the eigenvalues of
the matrix $L$ (in any representation of $\G$) are constants of
motion. When $L$ depends on a spectral parameter $\l$ the spectral
curve, i.e. the curve of equation det~$(L(\l)-\mu)=0$ in the
$(\l,\mu)$-plane, is similarly conserved. If $L(\l)$ depends
algebraically on $\l$, this is the algebraic equation of a compact
Riemann surface. Under quite general conditions
Semenov-Tian-Shanskii~\cite{RSTS3} has shown that the solution of the
dynamical problem can be found using abelian functions defined on this
Riemann surface.  In any instance equation~(\ref{lax}) strongly
suggests a geometrical formulation of the dynamical problem as the
flow induced in $\G^*$ by Kirillov's symplectic structure~\cite{Ki}.

At this point it is important to notice that such a flow {\em is not
in general integrable}. In other words, the eigenvalues of $L$, while
conserved, and globally defined, do not allow for a solution of the
equations of motion. It was Liouville who first pointed out~\cite{Li}
that, in order to solve a system on a phase space of dimension $2n$,
it is necessary and sufficient to know $n$ integrals of motion,
globally defined, and {\em in involution} (i.e. the Poisson bracket of
any two integrals of motion is zero). Of course the actual hamiltonian
must belong to this set. Let us notice that, according to Darboux
theorem, it is {\em always} possible to complete the hamiltonian {\em
locally} by $(n-1)$ quantities in involution in order to get a
symplectic basis, so Liouville condition is non empty only when the
constants of motion are globally defined, as is the case for a Lax
spectrum.

It is remarkable that this Liouville formulation leads to a very
simple condition in the Lax setting, as first pointed out by Babelon
and Viallet~\cite{BV}. They have shown that the eigenvalues of $L$ are
in involution {\em if and only if there exists an $R$-matrix}, i.e. a
function $R$, globally defined on phase space with values in
$\G\otimes\G$, such that:
\begin{equation}
\{L\x L\}=[R,L\otimes {\bf 1}]-[R^\Pi,{\bf 1}\otimes L] \label{rm}
\end{equation}
In this equation, denoting $L=\sum_i L^i e_i$ where $L^i$ is a
dynamical quantity and $(e_i)$ a basis of $\G$ then $\{L\x L\}=
\sum_{i,j} \{L^i,L^j\}\, e_i\otimes e_j$. Similarly $R=\sum_{i,j}
R^{i,j}\, e_i\otimes e_j$ and $R^\Pi$ is the ``transposed'' quantity
$R^\Pi=\sum_{i,j} R^{j,i}\, e_i\otimes e_j$.  Finally $[R,L\otimes{\bf
1}]=\sum_{i,j,k} R^{i,j}L^k\, [e_i,e_k]\otimes e_j$, so that the
right-hand side is expressed in terms of the Lie algebra structure.

Moreover the Jacobi conditions on the Poisson bracket lead to a
constraint on the $R$-matrix. Under some simplifying hypothesis this
constraint may be brought under a form closely related (but not
identical) to the semi--classical limit of the Yang--Baxter equation.
We shall call this equation the classical Yang--Baxter equation. A
nice feature is that the classical Yang--Baxter equation becomes
expressed entirely in terms of the Lie algebra structure of $\G$. This
opens the way to the study of the solutions of this equation (in the
spirit of the Belavin--Drinfel'd analysis~\cite{BD}) in order to
partially classify integrable systems.

It appears by looking at examples that the available $R$-matrices
either do not involve a spectral parameter, or fall into various
classes with respect to it. As a matter of fact, antisymmetric
$R$-matrices (i.e such that $R^\Pi=-R$) exactly obey
Belavin--Drinfeld's analysis and may be classified into rational,
trigonometric, and elliptic type.  Let us recall that rational
antisymmetric $R$-matrices are poorly understood (in contrast to
trigonometric, and elliptic ones). A simple result is that one--pole
matrices for a simple Lie algebra ${\bf g}$ are of the form:
\begin{equation}
R={\Pi \over \l -\mu}. \label{runpole}
\end{equation}
Here, $\G$ is a loop algebra ${\bf g}\otimes {\bf C}[\l,\l^{-1}]$
and $R$ a constant element of $\G\otimes\G$,
viewed as a function $R(\l,\mu)$, and we are considering the
situation in which there is only one pole for $\mu=\l$. Finally $\Pi$
is the exchange operator for the underlying Lie algebra $\bf g$: $\Pi
(x\otimes y)=y\otimes x$.  The antisymmetry of $R$ results from the
symmetry of $\Pi$ and the antisymmetry of $(\l -\mu)$.

There is no systematic study of non--antisymmetric $R$-matrices but
examples show that two--poles non antisymmetric $R$-matrices are
involved quite generally. We do not know concrete examples involving
rational $R$-matrices with more than two poles, although it is easy to
construct such $R$-matrices by a meaning procedure~\cite{graded}. It should
be noted that trigonometric solutions can be obtained by an allowed
deformation of such $n$--poles solutions~\cite{ja}.

In the following we shall recall the derivation of the classical
Yang--Baxter equation, and then completely solve it for two-poles
$R$-matrices, under some natural hypothesis. This is a simplification
and elaboration of the argument of~\cite{AT3}.  We shall then interpret
the solution as implying that the Lax pair~(\ref{lax}) describes an
Adler-Kostant-Symes system, see~\cite{Ad,Ko,Sy}, on a twisted loop algebra.
This geometrical situation naturally leads to integrable systems, as
emphasized notably by Reiman and Semenov-Tian-Shanskii~\cite{RSTS2}. In
some sense this partially answers the classification problem for
integrable systems, showing that the above scheme appears necessarily
in a wide variety of situations. Unfortunately the cases with
$R$-matrices involving dynamical variables, or no spectral parameter
(notably for the Calogero model, see~\cite{JaTa}) are not covered by
such a geometrical formulation.

\section{The classical Yang--Baxter equation}

The central point of this paper being the study of the classical
Yang--Baxter equation, we shall recall here its derivation, which is
somewhat tricky. In order to get a neat proof it is convenient to
consider that the objects occurring in eq.~(\ref{rm}) are in fact
living in $T({\cal U}(\G))$, where ${\cal U}(\G)$ is the universal
algebra on the Lie algebra $\G$, containing notably ${\bf 1}, X$ for
$X\in \G$, $XY$ for $X$ and $Y\in \G$, such that $XY-YX=[X,Y]$ and so
on. Then $T({\cal U}(\G))$ is the tensorial algebra on ${\cal U}(\G)$
containing elements such as ${\bf 1}\otimes X,\;{\bf 1}\otimes XY,\;
X\otimes Y$ for $X,Y \in\G$ and so on. Elements in $T^n({\cal U}(\G))$
can be multiplied in the natural componentwise way, for example
$(X\otimes Y).(X^\prime\otimes Y^\prime)=(XX^\prime)\otimes
(YY^\prime)$ so that in particular $[X\otimes {\bf 1},X^\prime\otimes
Z]=[X,X^\prime]\otimes Z$.  The commutators occuring in eq.~(\ref{rm})
can be so interpreted.

It is then further convenient to introduce the following notations:
$$L_1=L\otimes{\bf 1}\otimes{\bf 1},\;L_2={\bf 1}\otimes L\otimes{\bf
1},\;L_3={\bf 1}\otimes {\bf 1}\otimes L,$$
$$R_{12}=\sum_{ij}\,R^{ij}\,e_i\otimes e_j\otimes {\bf 1},\;
R_{23}=\sum_{ij}\,R^{ij}\,{\bf 1}\otimes e_i\otimes e_j,\;
R_{31}=\sum_{ij}\,R^{ij}\,e_j\otimes {\bf 1}\otimes e_i$$ and the
similar objects $R_{21}=R_{12}^\Pi,\;R_{32}=R_{23}^\Pi,\;
R_{13}=R_{31}^\Pi$ obtained by $R^{ij}\to R^{ji}$ which are elements
of $T^3({\cal U}(\G))$.

Now, under the simplifying hypothesis that $R$ is a {\em constant
matrix, independent of the dynamical variables}, it is easy to
convince oneself that the iteration of equation~(\ref{rm}) leads to:
$$\{ \{ L \x L \} \x L\}= [R_{12} [R_{13},L_1]] - [R_{12}
[R_{31},L_3]] + [R_{21} [R_{32},L_3]] - [R_{21} [R_{23},L_2]]$$ Were
the $R$-matrix to contain dynamical terms, there would appear terms
with derivatives of $R$  in this equation. Moreover the circular
permutations involved in Jacobi identity for the Poisson bracket
reduce to the sum of circular permutations on indices $(1,2,3)$.

In so doing the term involving $L_1$ may be written: $$[R_{12}
[R_{13},L_1]] - [R_{13} [R_{12},L_1]] + [R_{32} [R_{13},L_1]] -
[R_{23} [R_{12},L_1]]$$ Noticing that Jacobi identity for the
commutator in $T^3({\cal U}(\G))$ may be used, the first and second
term produce $[[R_{12},R_{13}],L_1]$. Moreover
$[R_{32},L_1]=[R_{23},L_1]=0$ trivially so that the two other terms
can also be written as similar commutators, leading to the condition:
$$[\,[R_{12},R_{13}]+[R_{12},R_{23}]+[R_{32},R_{13}]\, , L_1]=0$$
Here, one introduces a second simplifying hypothesis: one assumes that
$L$ is general enough so that the only natural solution to this
compatibility condition is given by the classical Yang--Baxter
condition:
\begin{equation}
[R_{12},R_{13}]+[R_{12},R_{23}]+[R_{32},R_{13}]=0. \label{cyb}
\end{equation}

Some remarks are in order:
\begin{itemize}
\item Equation~(\ref{cyb}) implies the same equation with permutations
of indices as may be easily derived by inserting appropriate operators
$\Pi$, hence no new condition is obtained with the terms involving $L_2$
and $L_3$.
\item This equation is only valid as a constraint for a constant
$R$-matrix since otherwise terms with derivatives of $R$  occur
(particularly $\{ R,L \}$).
\item Eq.~(\ref{cyb}) is a sufficient condition for eq.~(\ref{rm})
to be consistent with the Jacobi identity, but the extent of its
necessity is not clear.
\item Finally this equation is closely related to the semi--classical
limit of the quantum Yang-Baxter equation, but not identical.
\end{itemize}

More precisely the compatibility equation for the quantum commutation
relations: $$ R \, . \, T_1 \otimes T_2 = T_2 \otimes T_1 \, . \, R$$
is known as the Yang--Baxter equation.  It reads:
$$R_{12}R_{13}R_{23}=R_{23}R_{13}R_{12} . $$ Considering the
$\hbar$--expansion: $R=1+\hbar r + \hbar^2 + \dots$ and collecting
terms up to order $\hbar^2$ in the above equation, one sees that $s$
terms cancel while $r$ terms lead to:
$$[r_{12},r_{13}]+[r_{12},r_{23}]+[r_{13},r_{23}]=0.$$ The difference
with eq.~(\ref{cyb}) is the occurence of $r_{23}$ and not $r_{32}$. As
a matter of fact, if one assumes an antisymmetric $R$-matrix, one gets
$R_{32}=-R_{23}$ and eq.~(\ref{cyb}) reduces exactly to the
semi--classical limit of the quantum Yang--Baxter equation. However,
there is no justification for such an hypothesis and concrete
mechanical examples lead to non--antisymmetric $R$-matrices~\cite{graded}.

\section {Two--poles solutions of the Yang--Baxter equation}

For many interesting integrable models the Lax matrix lives in a loop
algebra, i.e. may be written $L(\l)$ where $\l$ is a spectral parameter
and $L(\l)$ may be seen as a function with values in a simple Lie algebra $\G$.
Then eq.~(\ref{rm}) reads:
$$\{ L(\l)\x L(\mu) \} = [R(\l,\mu),L(\l)\otimes {\bf 1}] -
[R^\Pi(\mu,\l),{\bf 1}\otimes L(\mu)]$$
where the function $R(\l,\mu)$ has values in $\G \otimes \G$.

The symplectic structure of the dynamical system is characterized by
$R(\l,\mu)$ and notably by its pole structure. In particular the
simplest example of solution of eq.~(\ref{cyb}) is given by the
one--pole {\em antisymmetric} solution given by eq.~(\ref{runpole}).
This solution plays a central role in the general discussion of
antisymmetric solutions of eq.~(\ref{cyb}) by Belavin and
Drinfel'd~\cite{BD}.  Nevertheless, many interesting mechanical examples
require a non--antisymmetric $R$-matrix, and up to now nothing more
complicated than a two--pole solution of eq.~(\ref{cyb}) had to be
considered.  Accordingly we shall look for solutions of
eq.~(\ref{cyb}) of the form:
\begin{equation}
R(\l,\mu)={A\over \l -\mu}+{B\over \l +\mu} \label{rdp}
\end{equation}

Since a Lax matrix $L(\l)$ may be multiplied by $f(\l)$ for any
analytic $f$ and moreover $\l$ may be changed into $g(\l)$,
eq.~(\ref{rdp}) represents a general two--pole $R$-matrix, with $A,B
\in \G$.  In the spirit of Belavin and Drinfel'd analysis, we shall
assume that $\G$ is a simple Lie algebra, and obtain $A$ and $B$ such
that eq.~(\ref{cyb}) is satisfied. Then as we have shown in~\cite{AT4}
there is no allowed deformation of this solution by functions of $\l$
and $\mu$ up to the above mentioned freedom of redefinition.

The substitution of ansatz~(\ref{rdp}) in equation~(\ref{cyb}) leads
to the following condition:
$$\displaylines{
{[A_{12},A_{13}]\over (\l_1-\l_2)(\l_1-\l_3)}+
{[B_{12},B_{13}]\over (\l_1+\l_2)(\l_1+\l_3)}+
{[A_{12},B_{13}]\over (\l_1-\l_2)(\l_1+\l_3)}+ \hfill\cr
{[B_{12},A_{13}]\over (\l_1+\l_2)(\l_1-\l_3)}+
{[A_{12},A_{23}]\over (\l_1-\l_2)(\l_2-\l_3)}+
{[B_{12},B_{23}]\over (\l_1+\l_2)(\l_2+\l_3)}+ \hfill\cr
{[A_{12},B_{23}]\over (\l_1-\l_2)(\l_2+\l_3)}+
{[B_{12},A_{23}]\over (\l_1+\l_2)(\l_2-\l_3)}+
{[A_{32},A_{13}]\over (\l_3-\l_2)(\l_1-\l_3)}+ \hfill\cr
{[B_{32},B_{13}]\over (\l_3+\l_2)(\l_1+\l_3)}+
{[A_{32},B_{13}]\over (\l_3-\l_2)(\l_1+\l_3)}+
{[B_{32},A_{13}]\over (\l_3+\l_2)(\l_1-\l_3)}=0\hfill\cr}$$
Taking the pole in $(\l_1-\l_2)$ one gets:
$${[A_{12},A_{13}+A_{23}] \over \l_1-\l_3}+{
[A_{12},B_{13}+B_{23}] \over \l_1+\l_3}=0$$
which must be true for any $\l_1,\l_3$ whence:
$$[A_{12},A_{13}+A_{23}]=[A_{12},B_{13}+B_{23}]=0.$$
Similarly taking
the pole in $(\l_1+\l_3)$ one gets:
$$[B_{12},B_{13}-A_{23}]=[B_{12},A_{13}-B_{23}]=0.$$
Conversely if these 4 conditions are satisfied, they are also true
with permuted indices, so that the classical Yang--Baxter
equation is satisfied.

In order to analyze further these conditions it is highly convenient
to use a trick introduced for this purpose by Belavin and Drinfel'd,
known as {\em dualization}. Since $\G$ is a simple Lie algebra, it is
equipped with an essentially unique invariant scalar product denoted
$(\, , \,)$. Then to each element $A=u \otimes v$ spanning $\G \otimes \G$ we
can associate the endomorphism of $\G$: $$X \longrightarrow (v,X) u$$
We shall denote by the same letter $A$ this element of End$(\G)\simeq
\G \otimes \G^*$ since this association is an isomorphism. It enjoys
the nice property that the transposed endomorphism $^T \! A$ such that
$(AX,Y)=(X,^T \! AY)$ is associated to the operator $v\otimes u \in \G
\otimes \G$ which is the image of $u \otimes v$ by $\Pi$.

In order to obtain the dualized form of the above four equations, it
is convenient to proceed as follows: write $A= \sum_\alpha u_\alpha
\otimes v_\alpha$ with $u_\alpha, v_\alpha \in \G$ and for any $X,Y,Z
\in \G$ consider the scalar product:
$$\displaylines{
\left( X\otimes Y\otimes
Z,[A_{12},A_{13}+A_{23}]\right)= \hfill\cr
\sum_{\alpha,\beta} \, (X,[u_\alpha,u_\beta])(Y,v_\alpha)(Z,v_\beta)+
(X,u_\alpha)(Y,[v_\alpha,u_\beta])(Z,v_\beta)= \hfill\cr
(X,[AY,AZ])+(Y,[\, ^T \! AX,AZ])=(X,[AY,AZ]-A[Y,AZ]) \hfill\cr}$$
by using the invariance of the scalar product and the definition of
the transposition. Hence, $[A_{12},A_{13}+A_{23}]=0$
if and only if the dualized operator obeys $A[X,AY]=[AX,AY]$
as an element of End$(\G)$ for any $X,Y \in \G$.

By dualizing similarly the above four equations one gets finally:
\begin{equation}
A[X,AY]=[AX,AY] \quad; \quad A[X,BY]=[AX,BY] \label{eqa}
\end{equation}
\begin{equation}
B[X,AY]=-[BX,BY] \label{eqbb}
\end{equation}
\begin{equation}
B[X,BY]=-[BX,AY] \label{eqba}
\end{equation}

Assuming, as in the analysis of Belavin and Drinfel'd that the
dualized endomorphism $R(\l,\mu)$ is invertible at least for one
couple $(\l_0,\mu_0)$ one sees that any $Y \in \G$ can be written
under the form $A Y_1 + B Y_2$.  Then equations~(\ref{eqa}) are
equivalent to: $$A[X,Y]=[AX,Y]\qquad \forall X,Y \in \G.$$

Since $\G$ is simple and $A \neq 0$ this implies that $A$ is
invertible (since Ker$\,A$ is an ideal of $\G$) and similarly
Ker$\,(A-\l I)$	is an ideal hence there exists $\l_0$ such that
$A=\l_0 I$.  One can normalize so that $A=I$. Then eq.~(\ref{eqbb})
with $\sigma=-B\in {\rm End}\,(\G)$ reads $\sigma [X,Y]=[\sigma
X,\sigma Y]$, hence $\sigma$ is an automorphism of $\G$. Finally
equation~(\ref{eqba}) reads $\sigma[X,\sigma Y]=[\sigma X,Y]=[\sigma
X, \sigma^2 Y]$ hence $[\sigma X, Y-\sigma^2 Y]=0\quad \forall X,Y \in \G$.
But $\sigma X$ is arbitrary (noticing that Ker$\,(\sigma)$ is an ideal
of $\G$, hence $\sigma$ is bijective), so that with $Z=(\sigma^2 -1)Y$
one has ad$_Z = 0$.  Then for any $T \in \G, \; (Z,T) ={\rm Tr}\; {\rm
ad}_Z {\rm ad}_T =0$ hence $Z=0$ since $\G$ is simple and the Killing
form is not degenerate. Finally $\sigma^2 =1$ meaning that $\sigma$ is
an involutive automorphism of $\G$.

Finally, $R(\l,\mu)=1/(\l -\mu) - \sigma /(\l +\mu)$
or in the usual non dualized form:
\begin{equation}
R(\l,\mu)= {\Pi \over \l -\mu}-{\sigma \otimes {\bf 1}. \Pi
\over \l +\mu} \label{rsym}
\end{equation}
This is the general two--pole solution of the classical Yang--Baxter
equation under the above mentioned hypothesis.  A similar $n$--pole
solution can be constructed using an automorphism $\sigma$ such that
$\sigma^m =1$. It reads~\cite{graded}: $$R=\sum_{n=0}^{m-1} {\sigma^n
\otimes {\bf 1}.\Pi \over
\epsilon^n \l -\mu}\qquad \epsilon={\rm e}^{2 \pi i \over m}$$

\section{Geometrical interpretation}

Mechanical systems with the above $R$-matrices can be interpreted
geometrically in the framework of the Adler-Kostant-Symes
scheme~\cite{Ad,Ko,Sy}. We shall briefly sketch the relevant ideas. In the
present situation, one may identify $\G$ and $\G^*$, and introduce the
natural Kirillov's symplectic bracket on $\G$. For linear functions of
the form $f_X(Y)=(X,Y)\quad (X,Y \in \G)$ the value of the Poisson
bracket is $\{ f_X,f_Y \}=f_{[X,Y]}$.  This extends naturally to
products of linear functions, and finally to arbitrary functions on
$\G$. This Poisson bracket is degenerate since invariant functions on
$\G$ have a vanishing bracket with any other functions. One gets a
non--degenerate symplectic structure by restricting oneself to {\em
orbits} of the Lie action on $\G$. Now the idea is to interpret the
Lax equation of motion~(\ref{lax}) as describing the flow on such an
orbit according to an appropriate Lie structure.

As a matter of fact it is impossible to get enough information by
using just one Lie structure on $\G$.  The clever idea of Adler,
Kostant, Symes is to use the interplay between two Lie structures on
$\G$. Then one chooses dynamical systems parametrizing orbits of one
of the Lie structures, with the associated symplectic structure, while
integrable hamiltonians are given as invariants of the other Lie
structure, and may easily be shown to Poisson commute. A natural way
to produce such a situation is to take for $\G$ a loop algebra ${\bf
g}\otimes {\bf C}[\l,\l^{-1}]$ and for the second Lie algebra
structure on $\G$, impose the vanishing of brackets between positive
and negative powers of $\l$, carefully treating the zero--power.

It has been emphasized by Semenov-Tian-Shanskii~\cite{STS1}, that the classical
$R$-matrix appearing in eq.~(\ref{rm}) is simply a way to define a
second Lie algebra structure on $\G$, namely, for the dualized $R$:
\begin{equation}
[X,Y]_R=[X,RY]+[RX,Y] \label{rsem}
\end{equation}
so that the Lax equation~(\ref{lax}) is the natural equation of motion
for the corresponding AKS--scheme.  It should be remarked that this
works for a ``constant'' $R$-matrix, and that the condition for
$[~]_R$ to be a Lie bracket boils down to the classical Yang-Baxter
equation.

We shall now indicate the corresponding geometrical interpretation of
our two--pole $R$-matrix. According to eq.~(\ref{rsem}) it is
necessary to interpret the dualized $R$-matrix $R=1/(\l
-\mu)-\sigma/(\l+\mu)$.  The relation with positive and negative
powers in the loop algebra is provided by writing formally: $${1 \over
\l -\mu}={1\over 2}\sum_{n=0}^\infty \left( {\mu^n
\over\l^{n+1}}-{\l^n\over\mu^{n+1}}\right)\qquad {1 \over \l
+\mu}={1\over 2}\sum_{n=0}^\infty \left((-1)^n {\mu^n
\over\l^{n+1}}+{\l^n\over\mu^{n+1}}\right)$$ Terms such as
${\mu^n / \l^{n+1}}$ may be dualized by considering their action on
the loop algebra: $$\left({\mu^n\over\l^{n+1}},X(\mu)\right)=
\sum_k  \left({\mu^n\over\l^{n+1}},{X_k \over \mu^{k+1}}\right)
={X_k\over\l^{n+1}}$$ i.e. negative powers are reproduced while
positive ones are killed. So doing, on gets the completely dualized
form of eq.~(\ref{rsym}):
\begin{equation}
R=-{1+\hat{\sigma} \over 2}(P_+ -P_-) \label{rgeom}
\end{equation}
where $P_\pm$ are projectors on positive and negative powers, while
$\hat{\sigma}$ is the extension of $\sigma$ to the loop algebra:
$(\hat{\sigma})(X(\l))=
\sigma\, . \, X(-\l)$.

Here the projector $(1+\hat{\sigma} )/ 2$ restricts us to the
well--known twisted loop algebra~\cite{Kac} associated to an involutive
automorphism, and the $R$-matrix~(\ref{rsym}) identifies to the structure
introduced
notably by Reiman and Semenov-Tian-Shanskii~\cite{RSTS2}, i.e an AKS scheme
on a twisted algebra. We have just shown that this construction is the
more general compatible with the above mentioned hypothesis.

{\bf Acknowledgements} We thank warmly J. Avan and O. Babelon for many
discussions on the subject.

\end {document}